\documentclass[%
  reprint,
  superscriptaddress,
  preprintnumbers,
  amsmath,amssymb,
  aps,
]{revtex4-2}

\usepackage{natbib}
\usepackage{graphicx}
\usepackage{newpxtext}
\usepackage[osf]{mathpazo}
\usepackage{enumerate}
\usepackage[caption=false]{subfig}
\usepackage{color}

\begin{document}
\preprint{KEK-TH-2408}
\preprint{J-PARC-TH-0271}
\title{Analytic Map of Three-Channel S Matrix \\
-Generalized Uniformization and Mittag-Leffler Expansion-}
\date{\today}
\author{Wren A. Yamada}
\email{wren-phys@g.ecc.u-tokyo.ac.jp}
\affiliation{Department of Physics, Faculty of Science, University of Tokyo, 7-3-1 Hongo Bunkyo-ku Tokyo 113-0033, Japan}%
\affiliation{Theory Center, Institute of Particle and Nuclear Studies (IPNS), High Energy Accelerator Research Organization (KEK), 1-1 Oho, Tsukuba, Ibaraki, 205-0801, Japan}%

\author{Osamu Morimatsu}
\email{osamu.morimatsu@kek.jp}
\affiliation{Department of Physics, Faculty of Science, University of Tokyo, 7-3-1 Hongo Bunkyo-ku Tokyo 113-0033, Japan}%
\affiliation{Theory Center, Institute of Particle and Nuclear Studies (IPNS), High Energy Accelerator Research Organization (KEK), 1-1 Oho, Tsukuba, Ibaraki, 205-0801, Japan}%
\affiliation{Department of Particle and Nuclear Studies,
Graduate University for Advanced Studies (SOKENDAI),
1-1 Oho, Tsukuba, Ibaraki 305-0801, Japan}%

\author{Toru Sato}
\email{tsato@rcnp.osaka-u.ac.jp}
\affiliation{Research Center for Nuclear Physics (RCNP), Osaka University, Ibaraki, Osaka 567-0047, Japan}%
\begin{abstract}
We explore the analytic structure of the three-channel $S$ matrix by generalizing uniformization and making a single-valued map for the three-channel $S$ matrix.
First, by means of the inverse Jacobi’s elliptic function we construct a transformation from eight Riemann sheets of the center-of-mass energy squared complex plane onto a torus, on which the three-channel $S$ matrix is represented single-valued.
Secondly, we show that the Mittag-Leffler expansion, a pole expansion, of the three-channel scattering amplitude includes not only topologically trivial but also nontrivial contributions and is given by the Weierstrass zeta function.
Finally, we examine the obtained formula in the context of a simple three-channel model.
Taking a simple non-relativistic effective field theory with contact interaction for the $S=-2$,  $I=0$, $J^P = 0^+$, $\Lambda\Lambda-N\Xi-\Sigma\Sigma$ coupled-channel scattering,
we demonstrate that the scattering amplitude as a function of the uniformization variable is, in fact,  given by the Mittag-Leffler expansion with the Weierstrass zeta function
and that it is dominated by contributions from neighboring poles.
\end{abstract}
\maketitle
The spectrum of excited states and their structure
  are the central issues in the study of interacting quantum systems
  such as nuclear, hadron, atomic and molecular systems.
  Those excited states show up as resonances embedded in the continuum
  spectrum.  Extraction of resonance information from the continuum spectrum is one of the key subjects.
\par
 In hadron physics, many candidates of exotic hadrons have recently been found near
 the threshold of new hadronic channels \cite{Guo:2017jvc,karliner}.
However, most of the existing analyses of the spectra for these signals are unsatisfactory.
For instance some assume the Breit-Wigner formula near the threshold, which cannot be justified,
and/or an arbitrary background, while some others depend on particular models.
\par
Very recently, a new method, the uniformized Mittag-Leffler expansion, has been proposed in Ref.\,\cite{PhysRevC.102.055201}
for extracting resonance information from the continuum spectrum.
The method is theoretically well-grounded, model independent, simple and useful.
The idea of the method is to find a variable in terms of which the $S$ matrix is single-valued (uniformization \cite{cohn2014conformal,Newton:book})
and to express the $S$ matrix as  a sum of pole terms (Mittag-Leffler expansion \cite{arfken2013mathematical,Nussenzveig:1972tcd}).
This is possible because the $S$ matrix is a meromorphic function owing to uniformization,
which then can be expressed as a sum of pole terms by the Mittag-Leffler theorem.
Therefore, it is crucial to find an appropriate uniformization variable which matches the analytic structure of the $S$ matrix.
\par
For the single-channel $S$ matrix, the center-of-mass momentum serves a role as the uniformization variable
,and the Mittag-Leffler expansion has already been studied in Ref.\,\cite{HUMBLET1961529,RAMIREZJIMENEZ201818}.
\par
For the two-channel $S$ matrix, a uniformization variable was introduced in Ref.\,\cite{KATO1965130,Newton:book}
and the uniformized Mittag-Leffler expansion has been examined in the past few years \cite{PhysRevC.102.055201,Yamada:2021cjo,PhysRevD.105.014034}.
It was applied to a model theory and shown to work in Ref.\, \cite{PhysRevC.102.055201}.
Then, it was applied to actual experimental data and was found to be very successful in extracting resonance
  information \cite{Yamada:2021cjo}.
Subsequently, the method was shown to be especially useful to understand the spectrum in the near-threshold region \cite{PhysRevD.105.014034}.
\par
Clearly, an extension of the method to three-channel reactions
is an important step for the analysis of multi-channel scattering.
Up to now, there only exist some works in which an approximated local uniformization has been employed for the analysis of three-channel scattering \cite{Krupa:1995fc,Surovtsev:2011yg,PhysRevA.8.754}.
It has been pointed out that the three-channel $S$ matrix can be
topologically mapped on a torus \cite{Newton:book,WEIDENMULLER196460},
however, as far as we know, an explicit formula of the uniformization variable for the three-channel $S$ matrix is yet to be known.

The purpose of the present paper is three-fold.
The first is to generalize uniformization.
For the first time, we explicitly show a uniformization variable for the three-channel $S$ matrix by means of the inverse Jacobi’s elliptic function.
The second is to obtain an explicit expression of the Mittag-Leffler expansion for the three-channel scattering amplitude taking account of both topologically trivial and nontrivial contributions.
The resulting expression is given by the Weierstrass zeta function.
The third is to demonstrate the validity of the obtained results in the context of a simple three-channel model.

Throughout this paper, we assume that the singularities of the three-channel $S$ matrix are poles and the three right-hand cuts starting from each threshold.
We do not consider left-hand cuts or other singularities.
\par
First, we generalize uniformization to the three-channel $S$ matrix.
For the explanation to be self-contained, we briefly summarize uniformization in the two-channel case.
\par
\begin{figure}[htbp]
    \includegraphics[width=\linewidth]{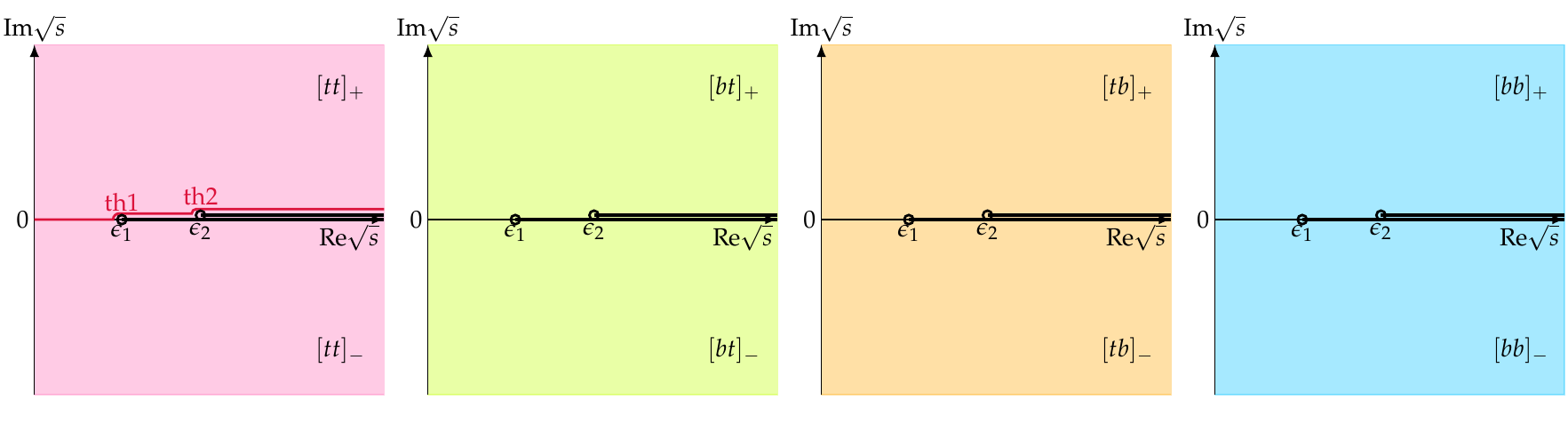}
    \caption{Four-sheeted complex $\sqrt{s}$-plane for two-channel S-matrix specified by a set of complex channel momenta as [$tt$],  [$tb$], [$bt$] and [$bb$],
where e.g.\ [$tb$]$_+$ means ${\rm Im}\, q_1 > 0$, ${\rm Im}\, q_2 < 0$ and ${\rm Im} \sqrt{s} >0$.
}
\label{Fig:Riemann_2_s}
\end{figure}

As a function of the center-of-mass energy, $\sqrt{s}$, the Riemann surface of a two-channel $S$ matrix is a four-sheeted complex plane with two branch points at $\sqrt{s}=\varepsilon_1$ and $\varepsilon_2$ ($\varepsilon_1 < \varepsilon_2$), as shown in Fig.\,\ref{Fig:Riemann_2_s}.
where $\varepsilon_i=M_i + M_i'$ is the threshold energy, $M_i$ and $M'_i$ are the masses of particles and $q_i=\sqrt{s-\varepsilon_i^2}$ is the \lq\lq momentum'' in the  $i$-th channel.
\par
We define the two-channel uniformization variable \cite{KATO1965130,Newton:book} $z$ by
\begin{align}
  z = \frac{ q_1 + q_2 }{\Delta_{12}},
\label{Eq:1}
\end{align}
where $\Delta_{12} = \sqrt{\varepsilon_2^2-\varepsilon_1^2}$.
Inversely, $q_1$ and $q_2$ are given as single-valued functions of $z$ by
\begin{align}
\begin{split}
  q_1 = \frac{\Delta_{12}}{2} \left(z + \frac{1}{z}\right), \quad
  q_2 = \frac{\Delta_{12}}{2} \left(z - \frac{1}{z}\right).
\end{split}
\label{Eq:2}
\end{align}
\par
\begin{figure}[ht]
  \begin{tabular}{cc}
    \begin{minipage}{0.5\hsize}
      \includegraphics[width=\linewidth]{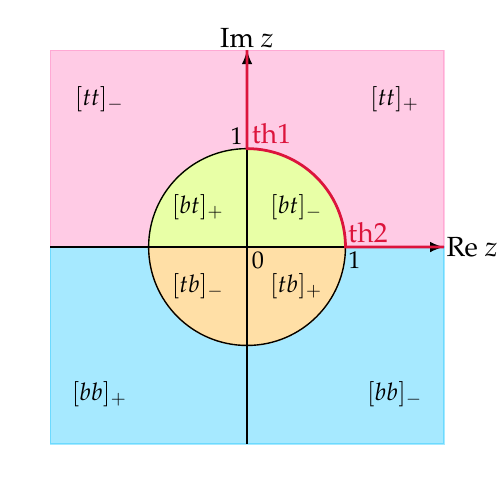}
    \end{minipage}
    \begin{minipage}{0.4\hsize}
      \includegraphics[width=\linewidth]{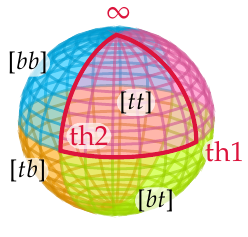}
    \end{minipage}
  \end{tabular}
  \caption{One-sheeted complex $z$-plane or a sphere for two-channel $S$ matrix.}
  \label{Fig:Riemann_2_z}
\end{figure}

Thus, the four Riemann sheets of the complex $\sqrt{s}$-plane are mapped onto one Riemann sheet of the complex $z$-plane or equivalently on a sphere as shown in Fig.\,\ref{Fig:Riemann_2_z}.
\par
\begin{figure}[h!]
    \includegraphics[width=\linewidth]{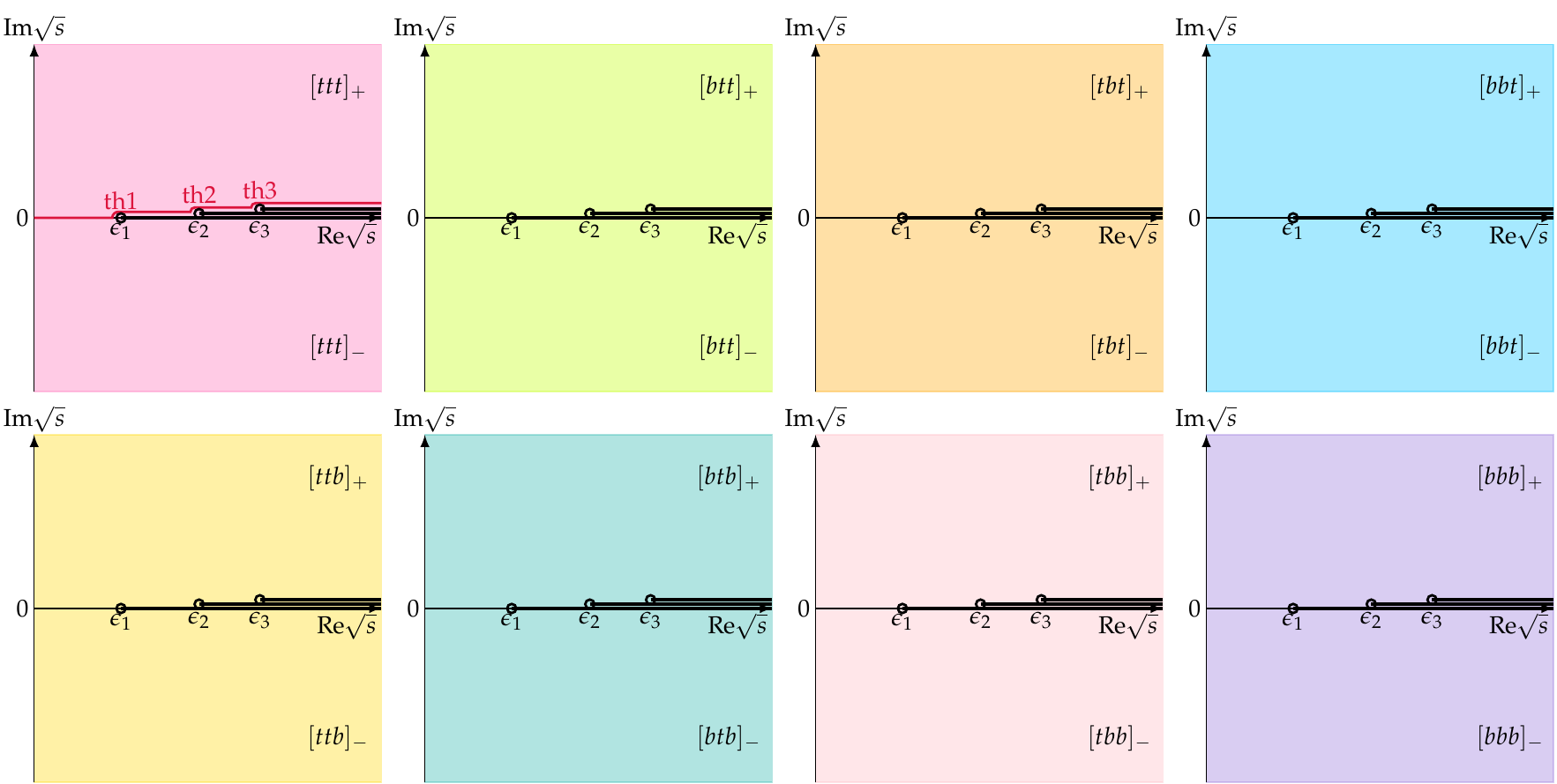}
    \caption{Eight-sheeted complex $\sqrt{s}$-plane for three-channel $S$ matrix specified by a set of complex channel momenta as [$ttt$],  [$ttb$], [$tbt$], [$tbb$], [$btt$],  [$btb$], [$bbt$] and [$bbb$], where e.g.\ [$ttb$]$_+$ means ${\rm Im}\, q_1 > 0$, ${\rm Im}\, q_2 > 0$, ${\rm Im}\, q_3 < 0$ and ${\rm Im} \sqrt{s} >0$.
}
  \label{Fig:Riemann_3_s}
\end{figure}

As a function of $\sqrt{s}$, the Riemann surface for a three-channel $S$ matrix is an eight-sheeted complex plane with three branch points at $\sqrt{s}=\varepsilon_1$, $\varepsilon_2$ and $\varepsilon_3$ ($\varepsilon_1 < \varepsilon_2 < \varepsilon_3$), as shown in Fig.\,\ref{Fig:Riemann_3_s}.
\par
The uniformization of the three-channel $S$ matrix is carried out in two steps.
We uniformize the channels 1 and 2 in the first step and the channel 3 in the second step.
For later use we define $\Delta_{12}$,  $\Delta_{13}$ and $\Delta_{23}$ by
$\Delta_{12} = \sqrt{\varepsilon_2^2-\varepsilon_1^2}$, 
$\Delta_{13} = \sqrt{\varepsilon_3^2-\varepsilon_1^2}$ and
$\Delta_{23} = \sqrt{\varepsilon_3^2-\varepsilon_2^2}$.
\par
The first step is the same as the uniformization of the two-channel $S$ matrix for the channels 1 and 2.
We define $z_{12}$ by
\begin{align}
	  z_{12} = \frac{ q_1 + q_2 }{\Delta_{12}},
\label{Eq:3}
\end{align}
Inversely, $q_1$ and $q_2$ are given as single-valued functions of $z_{12}$ by
\begin{align}
  q_1 = \frac{\Delta_{12}}{2} \left(z_{12} + \frac{1}{z_{12}}\right), \quad
  q_2 = \frac{\Delta_{12}}{2} \left(z_{12} - \frac{1}{z_{12}}\right).
\label{Eq:4}
\end{align}
$q_3$ is given by
\begin{align}
  q_3 = \frac{\Delta_{12}}{2}z_{12}\sqrt{\left(1-\frac{\gamma^2}{z_{12}^2}\right)\left(1-\frac{1}{\gamma^2z_{12}^2}\right)},
\label{Eq:5}
\end{align}
where
$\gamma = \frac{\Delta_{13} + \Delta_{23}}{\Delta_{12}}$.
$q_3$ is a double-valued function of $z_{12}$ with four branch points,
$
  z_{12} = \frac{\pm\Delta_{13} \pm\Delta_{23}}{\Delta_{12}} = \pm \gamma, \pm \frac{1}{\gamma}.
$
\par
\begin{figure}[h!]
      \includegraphics[width=\linewidth]{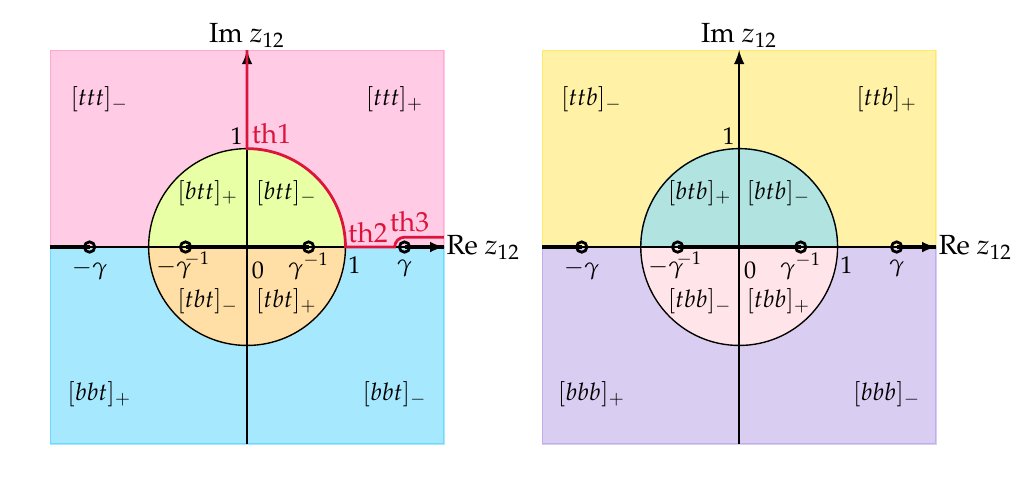}
\caption{Two-sheeted complex $z_{12}$-plane with four branch points of channel-3 for three-channel $S$ matrix.}
\label{Fig:Riemann_3_z12}
\end{figure}

Thus, in the first step, eight Riemann sheets of the complex $\sqrt{s}$-plane are mapped onto two Riemann sheets of the complex $z_{12}$-plane with four branch points as shown in Fig.\,\ref{Fig:Riemann_3_z12}.
\par
The second step is to uniformize the channel 3.
It is known that two Riemann sheets with four branch points is homeomorphic to a torus \cite{2019:otogi, cohn2014conformal}.
The map from the former with the coordinate $v$ to the latter with the coordinate $u$ is given by an elliptic integral.
In particular, when $v=\pm 1$ and $\pm 1/k$ ($k \neq 0$, $\pm 1$) are four branch points, the transformation is given by
\begin{align}
   u={\rm sn}^{-1}(v,k),
\label{Eq:6}
\end{align}
where ${\rm sn}^{-1}(v,k)$ is the inverse Jacobi's elliptic function \cite{2019:otogi, whittaker1928course},
\begin{align}
   {\rm sn}^{-1}(v,k) = \int_0^v \frac{dv}{\sqrt{(1-v^2)(1-k^2v^2)}}.
\label{Eq:7}
\end{align}
Inversely, $v$ is given as a single-valued function of $u$ by
\begin{align}
   v={\rm sn}(u,k).
\label{Eq:8}
\end{align}
\par
The Jacobi's elliptic function is doubly periodic,
\begin{align}
  {\rm sn}(u+2\omega_1,k) = {\rm sn}(u+2\omega_2,k) = {\rm sn}(u,k),
\label{Eq:9}
\end{align}
where
\begin{align}
\begin{split}
  2\omega_1 &= 4K(k) = 4\int_0^1 \frac{dv}{\sqrt{(1-v^2)(1-k^2v^2)}} \\
  2\omega_2 &= 2iK'(k) = 2i\int_0^1 \frac{dv}{\sqrt{(1-v^2)(1-(1-k^2)v^2)}}
\end{split}
.
\label{Eq:10}
\end{align}
Therefore, the image of the map, Eq.\,(\ref{Eq:6}) is a torus with two periods, $2\omega_1  = 4K(k)$ and $2\omega_2 = 2iK'(k)$.

\begin{figure}[htbp]
      \includegraphics[width=0.5\linewidth]{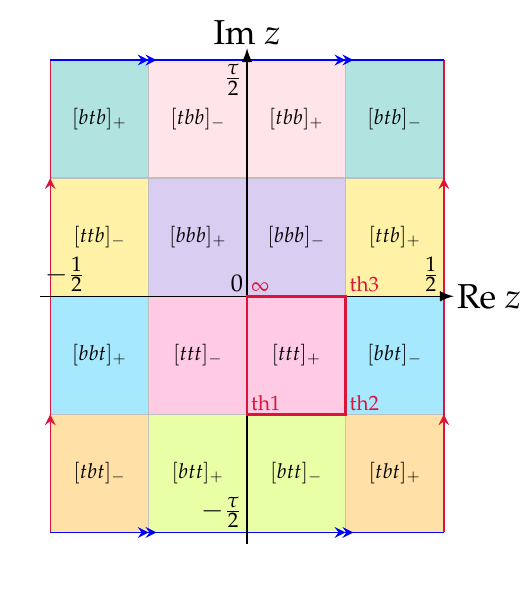}
      \raisebox{1.35cm}{\includegraphics[width=0.45\linewidth]{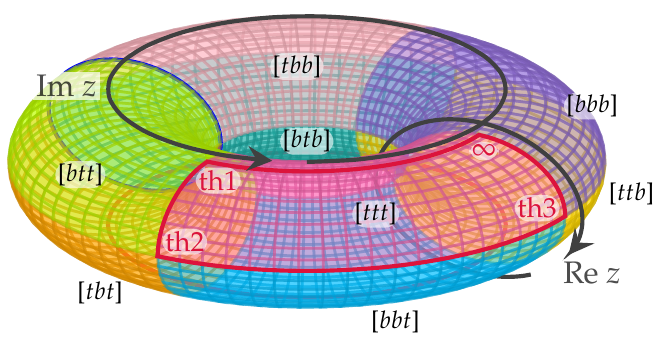}}
\caption{One-sheeted complex $z$-rectangle or a torus for three-channel $S$ matrix.}
\label{Fig:Riemann_3_z}
\end{figure}

Combining the first and second steps by taking $k=1/\gamma^2$,  $v=\gamma/z_{12}$, $u=4K(1/\gamma^2)z$,
we define the three-channel uniformization variable $z$ as
\begin{align}
  z = \frac{1}{4K(1/\gamma^2)} {\rm sn}^{-1}(\gamma/z_{12},1/\gamma^2),
\label{Eq:11}
\end{align}
where $z_{12}$ is given by Eq.\,(\ref{Eq:3}).
Inversely, $z_{12}$ is given as a single-valued function of $z$ by
\begin{align}
  z_{12}& = \frac{\gamma}{{\rm sn}(4K(1/\gamma^2)z,1/\gamma^2)}.
\label{Eq:12}
\end{align}
Therefore $q_1$, $q_2$ and $q_3$ are given as single-valued functions of $z$ by
\begin{align}
\begin{split}
  q_1&=\frac{\Delta_{12}}{2}\left(\frac{{\rm sn}(4K(1/\gamma^2)z,1/\gamma^2)}{\gamma}+\frac{\gamma}{{\rm sn}(4K(1/\gamma^2)z,1/\gamma^2)}\right) \\
  q_2&=\frac{\Delta_{12}}{2}\left(\frac{{\rm sn}(4K(1/\gamma^2)z,1/\gamma^2)}{\gamma}-\frac{\gamma}{{\rm sn}(4K(1/\gamma^2)z,1/\gamma^2)}\right) \\
  q_3&=\frac{\Delta_{12}}{2}\frac{\gamma\,{\rm sn}'(4K(1/\gamma^2)z,1/\gamma^2)}{{\rm sn}(4K(1/\gamma^2)z,1/\gamma^2)}
\end{split}.
\label{Eq:13}
\end{align}
By combining the first and second steps, the eight Riemann sheets of the complex $\sqrt{s}$-plane are mapped onto a torus with periods, $1$ and $i\tau$, where $\tau=\frac{4K'(1/\gamma^2)}{2K(1/\gamma^2)}$, as shown in Fig.\,\ref{Fig:Riemann_3_z}.
\par
Next, we investigate the Mittag-Leffler expansion of the three-channel scattering amplitude, ${\cal A}$,
which is related to the $S$ matrix,  $S$, as ${\cal A}_{ij}=\frac{S_{ij}-\delta_{ij}}{2i\sqrt{k_ik_j}}$. 
\par
The Mittag-Leffler expansion of a meromorphic function on a complex plane, ${\cal F}(z)$, is given by (see e.g.\ \cite{arfken2013mathematical,Nussenzveig:1972tcd})
\begin{align}
\begin{split}
  {\cal F}(z) &= \sum_i \frac{r_i}{z-z_i} + \text{subtraction terms} \\
  &= z^n \sum_i \frac{r_i}{\left(z-z_i\right)z_i^n} + \sum_{k=0}^{n-1} \frac{{\cal F}^{(k)}(0)}{k!} z^k,
\end{split}
\label{Eq:14}
\end{align}
where $\{z_i\}$ and $\{r_i\}$ are positions and residues of the poles of ${\cal F}(z)$.
When the sum, $\sum_i \frac{r_i}{z-z_i}$, is divergent, we have to subtract terms of order $z^k$ ($k = 0, \dots , n-1$) until the $n$-times subtracted sum, $z^n \sum_i \frac{r_i}{\left(z-z_i\right)z_i^n}$, is convergent.
This is exactly the same as in the dispersion theory \cite{Nussenzveig:1972tcd}.
\par
As a function of the uniformization variable, $z$, the three-channel scattering amplitude, ${\cal A}(z)$, is defined on a torus and hence is an elliptic function, a doubly periodic meromorphic function.
Therefore, there exist not only topologically trivial but also nontrivial contributions from poles and
the Mittag-Leffler expansion becomes
\begin{align}
\begin{split}
  {\cal A}(z) =& \sum_i \left( \frac{r_i}{z - z_i } + {\sum_{m,n }}'\frac{r_i}{z - z_i - \Omega_{m,n}} \right) \\
  &+ \text{subtraction terms}.
\end{split}
\label{Eq:15}
\end{align}
$\sum_i$ is a sum in the fundamental period rectangle $(-0.5 < {\rm Re}\, z \le 0.5$, $-0.5\tau < {\rm Im}\, z \le 0.5\tau)$, ${\sum'}_{m,n }$ denotes a sum over all integers $m$ and $n$, excluding $m=n=0$,
and $\Omega_{m,n} = 2m \omega_1 + 2n \omega_2$.
$\{z_i\}$ and $\{r_i\}$ are positions and residues of the poles of ${\cal A}(z)$ in the fundamental period rectangle.
The residues, $\{r_i\}$, satisfy \cite{2019:otogi, cohn2014conformal, whittaker1928course}
\begin{align}
  \sum_i r_i = 0.
\label{Eq:16}
\end{align}
${\sum}'_{m,n} \frac{r_i}{z - z_i - \Omega_{m,n}}$ is linearly divergent and ${\cal A}(z)$ is given in the form of Eq.\,(\ref{Eq:14}) with $n=2$,
unless the sum in the fundamental period rectangle causes further divergences.
We rewrite it as
\begin{align}
  {\cal A}(z) &= \sum_i r_i \zeta(z-z_i) + C_0 + C_1 z,
\label{Eq:17}
\end{align}
where ${\zeta}(z)$ is the Weierstrass zeta function \cite{2019:otogi, whittaker1928course} defined by
\begin{align}
\begin{split}
{\zeta}(z) &= \frac{1}{z} + \sum_{m,n}{}' \left( \frac{1}{z-\Omega_{m,n}} + \frac{1}{\Omega_{m,n}} + \frac{z}{\Omega_{m,n}^2} \right) \\
&= \frac{1}{z} + \sum_{m,n}{}' \frac{z^2}{\left(z-\Omega_{m,n}\right)\Omega_{m,n}^2}.
\end{split}
\label{Eq:18}
\end{align}
The Weierstrass zeta function is odd,
$\zeta(-z) = \zeta(z)$,
and is quasi periodic,
$\zeta(z+2\omega_1) = \zeta(z) + 2\eta_1$ and
$\zeta(z+2\omega_2) = \zeta(z) + 2\eta_2$,
for constants $\eta_1$ and $\eta_2$.
$C_0$ is determined from the asymptotic condition that the scattering amplitude should vanish as $s \rightarrow \infty$,
i.e.\ ${\cal A}(0)=0$, as
\begin{align}
  C_0 &= - \sum_i r_i \zeta(-z_i) = \sum_i r_i \zeta(z_i),
\label{Eq:19}
\end{align}
where $\zeta(-z) = \zeta(z)$ is used.
While, $C_1$ is determined from the periodicity of ${\cal A}(z)$,
i.e.\ ${\cal A}(z+2\omega_1) = {\cal A}(z+2\omega_2) = {\cal A}(z)$ as
\begin{align}
  C_1 = 0,
\label{Eq:20}
\end{align}
where Eq.\,(\ref{Eq:16}), $\zeta(z+2\omega_1) = \zeta(z) + 2\eta_1$ and $\zeta(z+2\omega_2) = \zeta(z) + 2\eta_2$ are used.
From Eqs.\,(\ref{Eq:17}), (\ref{Eq:19}) and (\ref{Eq:20}), we finally obtain
\begin{align}
  {\cal A}(z) &= \sum_i r_i \left( \zeta(z-z_i) + \zeta(z_i) \right),
\label{Eq:21}
\end{align}
where $r_i \left( \zeta(z-z_i) + \zeta(z_i) \right)$ is identified as the contribution of the $i$-th pole to the three-channel scattering amplitude, ${\cal A}(z)$.
\par
The generalized uniformization, Eq.\,(\ref{Eq:11}) together with Eq.\,(\ref{Eq:3}), its inverse, Eq.\,(\ref{Eq:13}), and the Mittag-Leffler expansion, Eq.\,(\ref{Eq:21}), are the main results of the formal part of the present paper.
\par
If $z_i$ is a pole of ${\cal A}(z)$ with a residue $r_i$, so is $-z_i^*$ with a residue $-r_i^*$
because of the unitarity of the $S$ matrix, $\mathcal{S}(-q_1^*,-q_2^*,-q_3^*) =\mathcal{S}^*(q_1,q_2,q_3)$ or
$\mathcal{S}(-z^*) =\mathcal{S}^*(z)$.
Therefore, poles appear either as pairs symmetrically with respect to the imaginary $z$ axis with complex residues related to each other,
or independently on the axis, ${\rm Re}\, z = 0$ or $0.5$, with purely imaginary residues. 
This guarantees the property of the scattering amplitude, ${\rm Im}\, {\cal A}_{ii} = 0$ $(\sqrt{s} < \varepsilon_1)$ \cite{PhysRevC.102.055201}.
\par
Now, we examine the above results in the context of a simple model of the three-channel scattering.
We consider the $S=-2$,  $I=0$, $J^P = 0^+$, $\Lambda\Lambda-N\Xi-\Sigma\Sigma$ coupled-channel scattering, in which possible existence of the $H$ particle  \cite{PhysRevLett.38.195} has extensively been studied both theoretically and experimentally.
Hereafter, channels $\Lambda\Lambda$, $N\Xi$ and $\Sigma\Sigma$ are referred to as 1, 2 and 3, respectively.
We generalize a non-relativistic effective field theory for nucleon scattering \cite{Kaplan:1996xu} to the coupled three-channel scattering.
(In Ref.\,\cite{PhysRevC.94.065207}, a similar non-relativistic effective field theory  extended to the coupled three-channel scattering is employed.)
The leading-order effective Lagrangian is given by
\begin{align}
  {\cal L}= B^\dagger \left(i \partial_t + \hat M + \frac{\nabla^2}{2 \hat M} \right) B - \frac{1}{2}  [BB]^\dagger \left( \hat  C_S + \hat  C_T\sigma \cdot \sigma \right) [BB],
\end{align}
where
\begin{align}
  B =
  \begin{pmatrix}
    N \\
    \Lambda \\
    \Sigma \\
    \Xi \\
\end{pmatrix}
\quad \text{and} \quad
  \left[ B B \right] =
\begin{pmatrix}
  \Lambda \Lambda \\
  N \Xi \\
  \Sigma \Sigma
\end{pmatrix}.
\end{align}
Baryon scattering in the flavor-singlet ${}^1S_0$ channel only depends on $\hat  C_S$ and  $\hat  C_T$ in the linear combination $\hat C = \hat  C_S - 3 \hat  C_T$ and
the interaction is assumed to be only in the flavor-singlet channel as
\begin{align}
\hat C = C
\begin{pmatrix}
\frac{1}{8} & \frac{1}{4} & - \frac{\sqrt{3}}{8} \\
\frac{1}{4} & \frac{1}{2} & - \frac{\sqrt{3}}{4} \\
- \frac{\sqrt{3}}{8} & - \frac{\sqrt{3}}{4} & \frac{3}{8} \\
\end{pmatrix},
\end{align}
where $C$ is a coupling of dimension $(\text{mass})^{-2}$.
\par
The $3 \times 3$ scattering amplitude, ${\cal \hat A}$, is given by
\begin{align}
i{\cal \hat A}= -i \hat C \left(\hat 1- \hat G \hat C\right)^{-1},
\end{align}
which is rescaled from the previous definition as $\frac{4\pi}{\sqrt{M_iM_j}} {\cal A}_{ij} \rightarrow {\cal A}_{ij}$.
$\hat G$ is the diagonal $3 \times 3$ Green function,
\begin{align}
\hat G=
\begin{pmatrix}
G_{\Lambda\Lambda} &0 & 0 \\
0 & G_{N\Xi} & 0 \\
0 & 0 & G_{\Sigma\Sigma} \\
\end{pmatrix}
.
\end{align}
$G_i= \frac{-i\mu_ik_i}{2\pi}$ and $\mu_i=\frac{M_iM'_i}{M_i+M'_i}$ is the reduced mass in the channel, $i$.
$k_i$ is a non-relativistic momentum, $\sqrt{s} = \varepsilon_i + \frac{k_i^2}{2\mu_i}$, which is proportional to $q_i=\sqrt{s-\varepsilon_i^2}$ in the leading order of $q_i$ as $k_i = \sqrt{{\frac{\mu}{\varepsilon}}}q_i$. In the following discussion the differences of $k_i$ and $\sqrt{{\frac{\mu}{\varepsilon}}}q_i$ is ignored for simplicity.
Masses $M_N$, $M_\Lambda$, $M_\Sigma$  and $M_\Xi$ are taken to be average masses of charged baryons in PDG \cite{Zyla:2020zbs}. 
In this model the number of poles is four,
which is due to an extremely simple structure of the interaction.
This model is used in order to demonstrate validity of our formalism but not to describe realistic physics.
\par
\begin{figure*}[!htb]
\begin{tabular}{cc}
    \begin{minipage}{0.5\linewidth}
    \includegraphics[width=\linewidth]{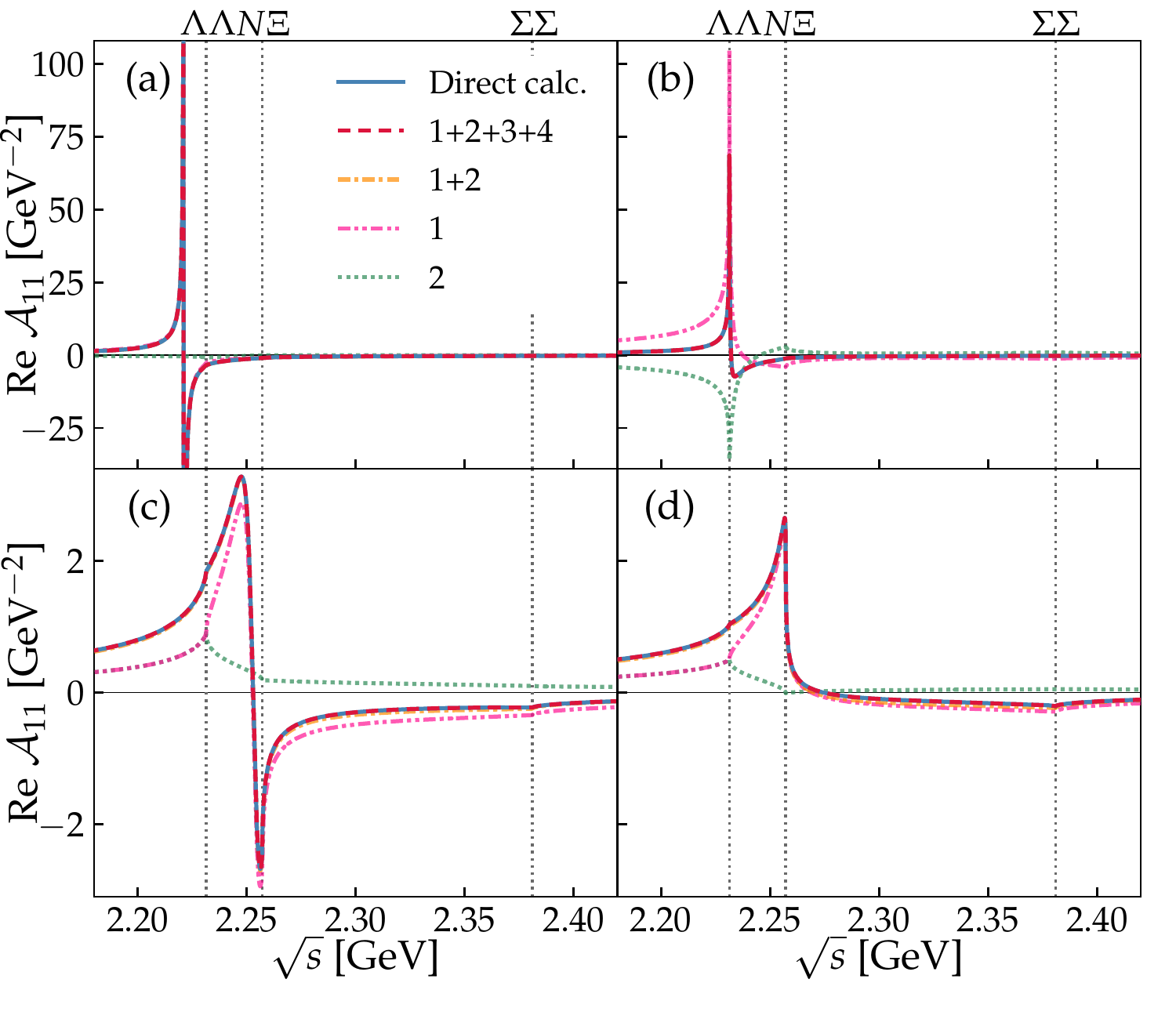}
    \end{minipage}
    \begin{minipage}{0.5\linewidth}
    \includegraphics[width=\linewidth]{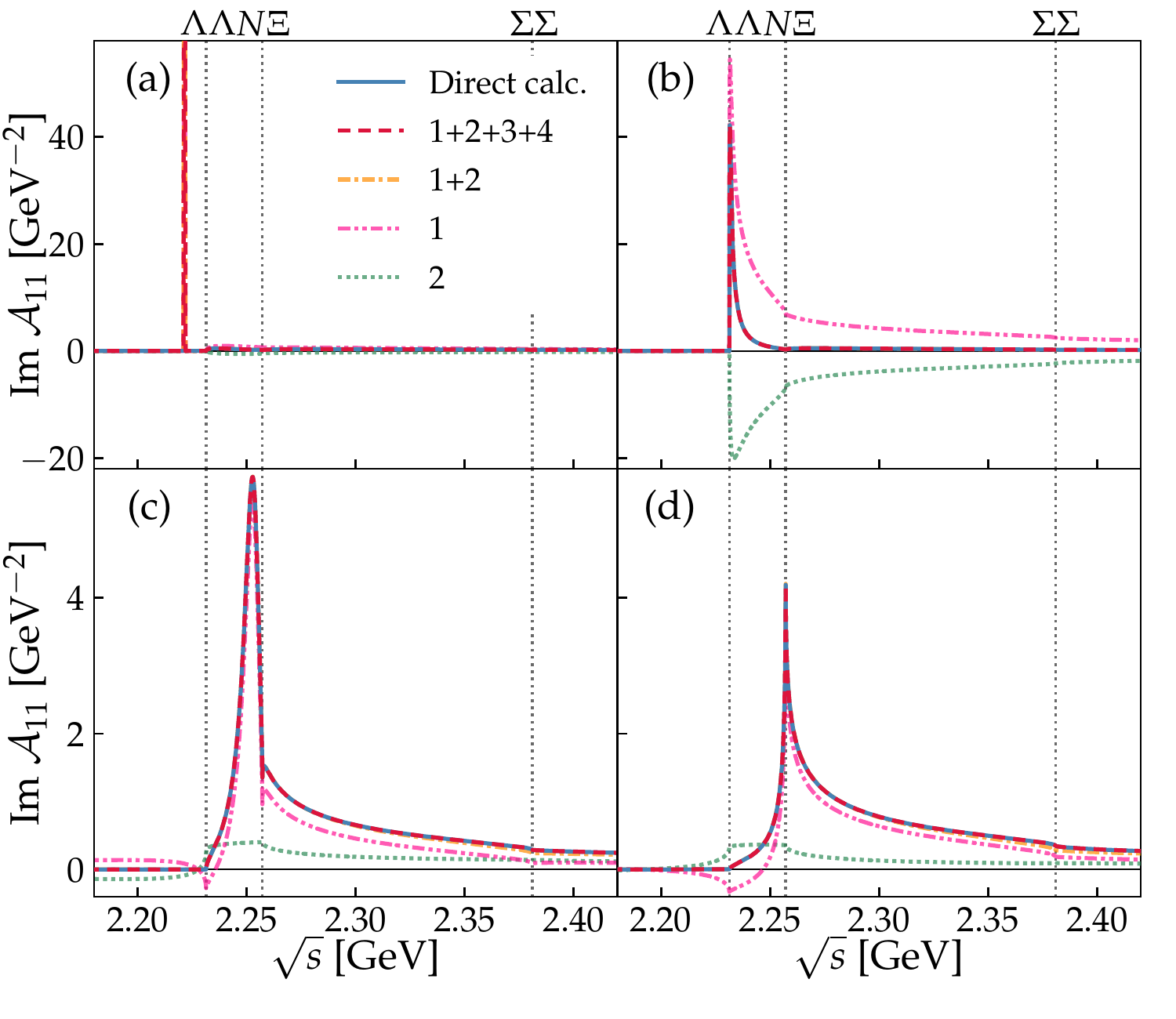}
    \end{minipage}
\end{tabular}
    \caption{Real and imaginary parts of the $\Lambda\Lambda\to\Lambda\Lambda$ elastic scattering amplitude, $\mathcal{A}_{11}$, for cases (a)-(d). The amplitudes of direct model calculation (blue), reconstructed via the uniformized Mittag-Leffler expansion with all four poles, 1+2+3+4 (red, dash-dotted), the contributions from pole 1, 2 and their sum, 1+2 (orange), respectively.}
\label{fig:respec}
\label{fig:spec}
\end{figure*}
\par
\begin{table*}[!htb]
  \begin{tabular*}{0.9\linewidth}{@{\extracolsep{\fill}}cccccc}
      \hline\hline
          &$C$ [GeV$^{-2}$]&Pole 1&Pole 2&Pole 3&Pole 4\\\hline
          &     &$-0.267i$&$-0.496i$&$0.5+0.043i$&$0.5-0.702i$\\
      (a) &$40.00$&$0.172i$&$-0.154i$&$-0.015i$&$-0.004i$\\
          &     &$2.221$ [$ttt$]&$2.200$ [$btt$]&$-1.802i$ [$ttb$]&$13.477i$ [$tbt$]\\\hline
          &     &$-0.371i$&$-0.398i$&$0.5+0.048i$&$0.5-0.700i$\\
      (b) &$45.60$&$1.750i$&$-1.727i$&$-0.018i$&$-0.005i$\\
          &     &$2.231$ [$btt$]&2.229 [$btt$]&$-1.252i$ [$ttb$]&$11.722i$ [$tbt$]\\\hline
          &     &$0.177-0.392i$&$-0.177-0.392i$&$0.5+0.060i$&$0.5-0.697i$\\
      (c) &$60.00$&$-0.215+0.018i$&$0.215+0.018i$&$-0.027i$&$-0.009i$\\
          &     &$2.253-0.005i$ [$btt$]&$2.253+0.005i$ [$btt$]&$0.907$ [$ttb$]&$8.657i$ [$tbt$]\\\hline
          &     &$0.271-0.402i$&$-0.271-0.402i$&$0.5+0.073i$&$0.5-0.691i$\\
      (d) &$80.00$&$-0.249+0.028i$&$0.249+0.028i$&$-0.038i$&$-0.017i$\\
          &     &$2.259+0.002i$ [$tbt$]&$2.259-0.002i$ [$tbt$]&$1.510$ [$ttb$]&$6.124i$ [$tbt$]\\\hline\hline
  \end{tabular*}
  \caption{Pole positions and residues of  the $\Lambda\Lambda\to\Lambda\Lambda$ elastic scattering amplitude, $\mathcal{A}_{11}$, for cases (a)-(d). The first and second rows are the pole positions, $z_i$, and residues, $r_i$, respectively,  on the torus ($-0.5<\text{Re}~z\leq 0.5$, $-0.5\tau<\text{Im}~z\leq 0.5\tau$).
  The third row is the complex center-of-mass energy of the pole, $\sqrt{s_i}$, in units of [GeV] and the complex Riemann sheet on which the pole is located.
  The threshold energies, $\varepsilon_1$, $\varepsilon_2$ and $\varepsilon_3$, are 2.231, 2.257 and 2.381 GeV, respectively.}
\label{tab:polpos}
\end{table*}
\begin{figure}[h]
    \centering
    \includegraphics[width=\linewidth]{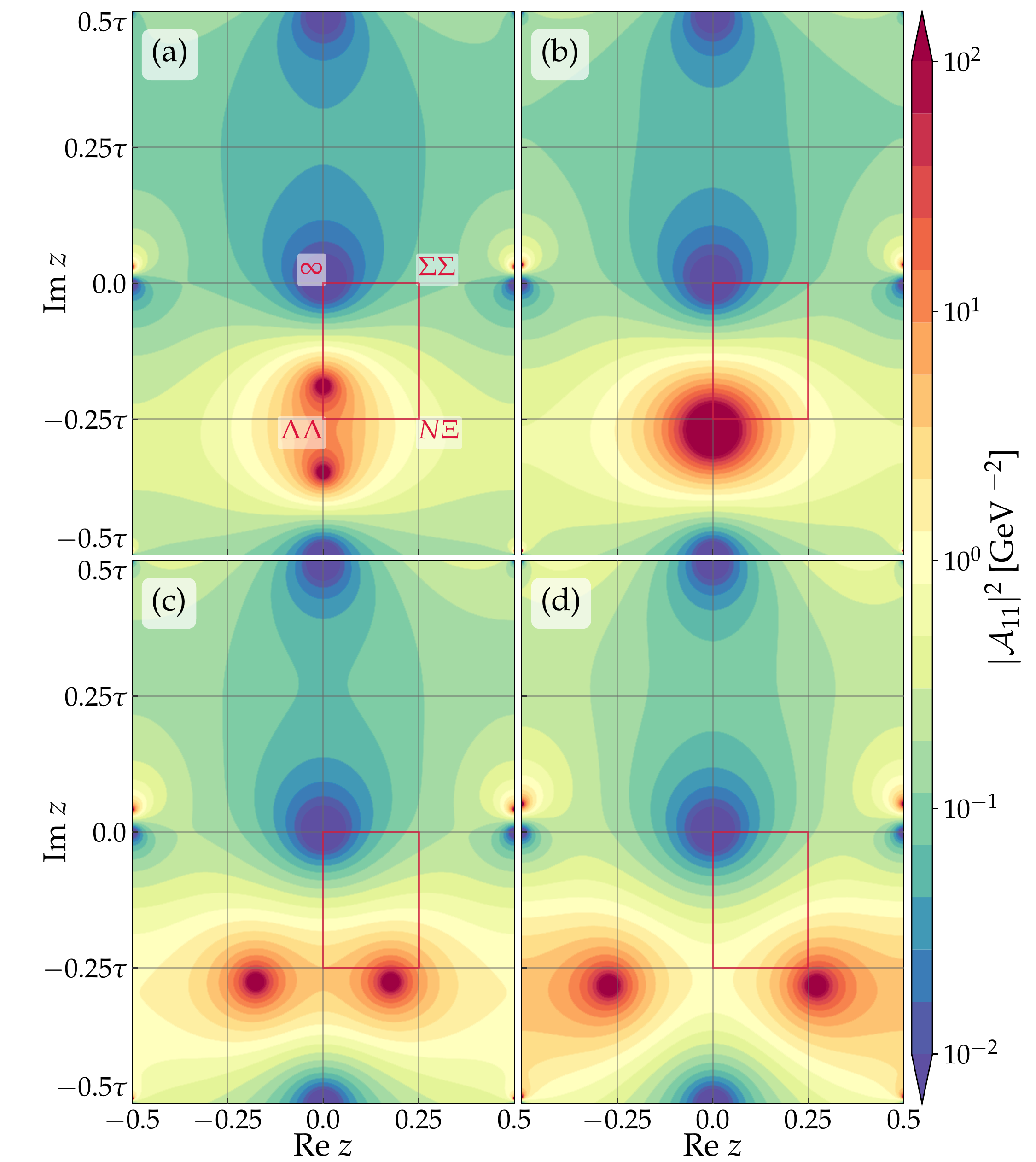}
    \caption{Contour plot of $|\mathcal{A}_{11}|^2$ on the torus ($-0.5<\text{Re}~z\leq 0.5$, $-0.5\tau<\text{Im}~z\leq 0.5\tau$) for cases (a)-(d). The red line corresponds to the physical domain. Labels $\Lambda\Lambda$, $N\Xi$, and $\Sigma\Sigma$ represent the corresponding thresholds, and $\infty$ corresponds to infinity point on the physical sheet.}\label{fig:trajec}
\end{figure}
Real and imaginary parts of the $\Lambda\Lambda\to\Lambda\Lambda$ elastic scattering amplitude, $\mathcal{A}_{11}$, are shown in Fig.\,\ref{fig:spec} for four different cases, (a)-(d), with different coupling, $C$, where we present the amplitudes of direct model calculation, reconstructed via the uniformized Mittag-Leffler expansion with all four poles, 1+2+3+4, the contributions from pole 1, 2 and their sum, 1+2, respectively.
Pole positions and residues of $\mathcal{A}_{11}$ are given in Tab.\,\ref{tab:polpos}.
\par
Sharp structures are observed below the $\Lambda\Lambda$ threshold (bound state) in case (a), at the $\Lambda\Lambda$ threshold (virtual state) in case (b), between the $\Lambda\Lambda$ and $N\Xi$ threshold (resonance) in case (c) and at the $N\Xi$ threshold (\lq\lq threshold cusp") in case (d).
The amplitudes of direct model calculation and reconstructed with all four poles, 1+2+3+4, perfectly coincide, which confirms our result, Eq.\,(\ref{Eq:21}).
The contribution from pole 1, which is nearest to the physical domain, gives the sharp structures of the amplitudes of direct model calculation.
The sum of contributions from poles 1 and 2 reproduces the amplitudes of direct model calculation in almost the entire physical domain,
which is due to an extremely simple nature of the model, i.e.\ the number of poles is four and only two of them are close to the physical domain, and will not be the case in a more realistic situation.
\par
Fig.\,\ref{fig:trajec} is the contour plot of $|\mathcal{A}_{11}|^2$ on the torus, a map of the three-channel $S$ matrix, for cases (a)-(d).
It can be observed that as the coupling, $C$, increases, pole 1 moves along the imaginary axis transitioning from a bound-state pole in case (a) on the [$ttt$]-sheet to a virtual-state pole in case (b) on the [$btt$]-sheet, then it leaves the imaginary axis in the positive real axis direction at a right angle and becomes a resonance pole on the [$btt$]-sheet in case (c), and finally a pole on the [$tbt$]-sheet, which causes \lq\lq threshold cusp" in case (d).
Pole 2 moves along the imaginary axis on the [$btt$]-sheet until it merges with pole 1. Then, it leaves the imaginary axis in the negative real axis direction and moves symmetrically to pole 1 with respect to the imaginary axis. 
Pole 3 and 4 hardly move.
From Fig.\,\ref{fig:trajec} together with Fig.\,\ref{fig:spec}, one can clearly observe that the effects of the poles show up as sharp structures on the scattering amplitude around the nearest physical energy region.
We would like to mention here that the use of the uniformization variable makes it extremely easy and transparent to locate the positions of poles.
When one trace poles on multi-sheeted complex $\sqrt{s}$ plane one has to move around different sheets.
\par
The above demonstration shows that the three-channel scattering amplitude is indeed given by the Mittag-Leffler expansion as a function of the uniformization variable, Eq.\,(\ref{Eq:21}),  and that the behavior of the scattering amplitude can intuitively be understood from the behavior of the poles by the use of the uniformization variable.
Also seen from the above demonstration is that a resonance pole smoothly transitions to a pole which causes \lq\lq threshold cusp", by the change of the interaction strength.
There is no essential difference between these two poles \cite{PhysRevD.105.014034}.
\par
To summarize, we generalized uniformization to the three-channel scattering by using the Jacobi's elliptic function and obtain the formula of the Mittag-Leffler expansion in terms of the Weierstrass's zeta function.
We also demonstrated in a simple model of the three-channel scattering that the obtained formula indeed holds.
\par
Having shown that our proposed method  is valid,
we are now planning to analyze actual experimental data, e.g.\,\cite{LHCb:2015yax,LHCb:2019kea,Belle:2018lws},
 by our method.
We hope that we will report the results of the analysis in a near future.
\par
We would like to thank Shun'ya Mizoguchi for the instruction on the construction of the Riemann surface. Without his help, the completion of this work would have been much harder.
We would also like to thank Akinobu Dote for bringing our attention to the $H$ particle as a possible application of the 3-channel uniformized Mittag-Leffler expansion. 
We would also like to thank Koichi Yazaki and the members of the discussion meeting held on the KEK Tokai campus, Yoshinori
Akaishi, Toru Harada, Fuminori Sakuma, Shoji Shinmura and Yasuhiro Yamaguchi.
\bibliography{3chan_unif}
\end{document}